# Symmetry Classification of Magnetic Orders using Oriented Spin Space Groups

Yuntian Liu[1,†], Xiaobing Chen[2,1,†], Yutong Yu[1], Jesús Etxebarria[3], J. Manuel Perez-Mato[3] and Qihang Liu[1,2,4,*]

*[1]State Key Laboratory of Quantum Functional Materials, Department of Physics, and Guangdong Basic Research Center of Excellence for Quantum Science, Southern University of Science and Technology (SUSTech), Shenzhen 518055, China*

*[2]Quantum Science Center of Guangdong–Hong Kong–Macao Greater Bay Area (Guangdong), Shenzhen 518045, China*

*[3]Faculty of Science and Technology, University of the Basque Country/Euskal Herriko, Unibertsitatea (UPV/EHU), Bilbao, Spain*

*[4]Guangdong Provincial Key Laboratory for Computational Science and Material Design, Southern University of Science and Technology, Shenzhen 518055, China*

[†]These authors contributed equally to this work.

[*]Email: liuqh@sustech.edu.cn

## Abstract

**Magnetism has witnessed remarkable progress in recent decades, largely driven by its potential for next-generation storage devices. However, the classification of magnetic orders, even for fundamental concepts such as ferromagnetism and antiferromagnetism, remains a topic of active evolution, particularly with the discovery of unconventional magnetic materials and advances in antiferromagnetic spintronics. Here, we present a unified classification of magnetic order utilizing the state-of-the-art spin space group (SSG) theory. Based on whether the net spin magnetization is constrained to zero by SSG, we systematically categorize magnetic orders into ferromagnetism (including ferrimagnetism) and antiferromagnetism. We further introduce an oriented SSG description, i.e., an SSG with a fixed magnetic orientation, thereby unifying the SSG and magnetic space group frameworks. This approach clearly reveals the symmetry-breaking pathway induced by spin-orbit coupling. The proposed group framework completes the intrinsic logic of magnetic symmetry and identifies a distinct magnetic phase, termed spin-orbit magnetism, in which the net spin magnetization is induced by spin-orbit coupling. Our work provides a comprehensive symmetry-based perspective for classifying magnetic order, offering fresh insights into unconventional magnets and broad applicability in spintronics and quantum material design.**

## *Introduction*

The established classification of magnetic order is fundamentally rooted in the dichotomy between ferromagnetism (FM) and antiferromagnetism (AFM). Recent advances in spintronics[1-6] have led to the discovery of novel magnetic phenomena and the emergence of materials with unconventional magnetism. These include geometrically complex spin textures, such as helimagnetism[7,8], as well as symmetry-protected magnetic states defined by specific physical responses, like altermagnetism[9] and kinetomagnetism[10]. While these new concepts are driving progress in spintronics, the absence of a rigorous and systematic classification framework has made it challenging to clarify the relationships between these unconventional magnetic phases and the conventional categories of FM and AFM[11].

Historically, the dichotomy of FM and AFM is based on the net spin magnetization $M_s$ within a unit cell. Specifically, AFM refers to an ordered spin geometry with zero net spin magnetization ($M_s = 0$) resulting from antiparallel spin alignment. In 1948, L. Néel proposed a refined definition of AFM under the condition $M_s = 0$, emphasizing that magnetic sublattices with opposite spins must be crystallographically equivalent[12]. In other words, the $M_s = 0$ condition, which applies to both collinear and noncollinear magnetic orders, must be enforced by certain symmetries. A classic example is the one-dimensional AFM chain, in which the two magnetic sublattices are related by translation symmetry. By contrast, magnets with inequivalent magnetic sublattices yet vanishing $M_s$, referred to as compensated ferrimagnets, typically display different behaviors in the magnetization versus temperature curve (*M-T*) and hysteresis loops compared to antiferromagnets, resembling ferromagnets in several aspects[13,14].

The key question is: how to mathematically rationalize the dichotomy of FM and AFM? To date, the symmetry of magnetic materials has predominantly been described within the framework of magnetic space groups (MSGs)[15-17]. The MSG approach has been profoundly successful and remains a cornerstone of magnetic crystallography. However, because MSGs by definition[18] restrict the rotation in spin space to be identical to that in real space (thereby implicitly incorporating spin-orbit coupling, SOC), they do not fully capture the spin geometries, which reflect the crystallographic feature driven by the isotropic spin exchange but are not robust against SOC. Fig. 1**a** shows three magnetic structures that can be easily categorized as collinear FM, collinear AFM, and coplanar AFM by bare eyes phenomenologically; however, they share the same MSG $Cm'cm'$. As another example, Fig. 1**b** considers three magnetic configurations of MnTe that share the same AFM order but differ in Néel vector orientation, resulting in three distinct MSGs. Therefore, while the different MSGs successfully capture the possible SOC response, e.g., the anomalous Hall effect (AHE), they inherently describe different physical aspects than the common crystallographic characteristics of the collinear AFM geometry driven by the same exchange interactions.

To provide a complementary perspective, a symmetry framework known as the spin space group (SSG)[19-23], which combines separate spin and spatial operations, can be used to characterize the correlations present in spin geometry arising from isotropic spin exchange. Using SSGs, we will show in this work that a stringent classification of the FM-AFM dichotomy can be established by the condition of having an SSG-enforced null net spin magnetization ($M_s = 0$). This mathematical condition can be taken as a systematic foundation for discussing "new types of magnetism".

Furthermore, to bridge these different methodologies, we introduce the concept of the "oriented SSG" (OSSG). The OSSG encompasses both the SSG and MSG of a structure within a single framework, treating them as a group-subgroup pair to provide a comprehensive, combined description of magnetic materials. We show that the FM-AFM dichotomy, combined with the condition of MSG-enforced null magnetization ($M = 0$), naturally underlies a class of unconventional magnetism, which can be termed spin-orbit magnetism (SOM) because the net spin magnetization is induced by SOC as a consequence of its symmetry breaking. By symmetry analysis and first-principles calculations, we identify 224 SOM candidates from the MAGNDATA database[24,25] and reveal their unique magnetization mechanisms, for instance, orbital moments arising from lower-order SOC terms compared with spin moments, implying potential for future devices with vanishing magnetization.

### ***FM and AFM dichotomy***

The definition by L. Néel[12] phrases the description of AFM order as a crystallographic problem, and it can indeed be rigorously defined within the symmetry framework of SSGs. Specifically, if the SSG of a given magnetic structure constrains its net spin magnetization to be zero, it is then classified as AFM; otherwise, it is classified as FM or ferrimagnetism. Furthermore, any nonzero $M_s$ not allowed by the SSG, but allowed by the corresponding MSG, must necessarily be a SOC effect and generally weak, compared with the arrangement with $M_s = 0$ permitted by the SSG.

The general group structure of an SSG consists of $\{g_s||g_l|\tau\}$ operations ($g_s$ is a spin-space operation including spin rotation and time-reversal symmetry $T$; $g_l$ is a real-space operation including spatial rotation and space-inversion $P$; and $\tau$ stands for a translation). Since the spin magnetization is solely restricted by $g_s$, we thus define the spin-space point group $P_{spin}$ by mapping $\{g_s||g_l|\tau\}$ operations to $g_s$. $P_{spin}$ encodes all the symmetry information of the spin-space, and can thus be utilized to classify AFM/FM by its polarity. Specifically, a non-polar $P_{spin}$ forbids all the components of the net spin magnetization, leading to SSG-enforced $M_s = 0$. In contrast, a polar $P_{spin}$ indicates that the SSG symmetry permits a net spin magnetization along a certain orientation. Therefore, FM can be easily identified by $P_{spin}$ being a subgroup of $\infty m$, while AFM is the opposite. We further categorize two classes of FM with polar $P_{spin}$ and four classes of AFM with non-polar $P_{spin}$, as shown in Extended Data Table I. Specifically, for collinear magnets, since $P_{spin}$ of a collinear SSG simultaneously has a mirror plane $m$ containing the collinear axis, $P_{spin}$ for collinear FM and collinear AFM must be $\infty m$ and $\infty/mm$, respectively.

We point out that the geometric features of a given magnetic structure, including its FM or AFM character and its collinearity or noncollinearity, can be ultimately classified and encoded in the International Symbol of its SSG (here we adopt the Chen-Liu symbol from Ref.[21]). Taking the magnetic structures in Fig. 1**a** as examples, for $LaMnSi_2$ (SSG: $C^1m^1c^1m^{\infty m}1$), its spin-only group $^{\infty m}1$ indicates the collinearity, and the $P_{spin}$ is the polar $\infty m$ that permits nonzero $M_s$ along the collinear axis. Consequently, it is categorized as a collinear FM. Similarly, the $P_{spin}$ of $CaIrO_3$ (SSG:

$C^{1}m^{-1}c^{-1}m^{\infty m}1$) and $Mn_3Ge$ (SSG: $P^{3^1}6_3/^{1}m^{2}m^{2}c^{m}1$) are $\infty/mm$ and $-62m$, respectively, both of which are non-polar, indicating AFM orderings. Furthermore, the spin-only group $^{\infty m}1$ for $CaIrO_3$ and $^{m}1$ for $Mn_3Ge$ denote their collinear and coplanar character, respectively.

In addition, SSGs also provide a rigorous symmetry description for complex AFM geometries beyond MSGs, such as helimagnets[7,8] and multiple-$q$ magnets[26-28], which are phenomenologically characterized by one or more propagation vectors $q$. Specifically, these geometries can be further classified by a subgroup of their SSG, named spin translational group $T_{spin}$. $T_{spin}$ consists of operations combining a pure spin-space operation and a fractional translation $\{g_s||1|\tau\}$. Thus, it directly describes the periodic arrangement of the spin relative orientations. Based on the order and the cyclic nature of $T_{spin}$, we classify AFM geometries into four distinct categories, ranging from typical Néel-type to multi-$q$ AFM, and present representative materials for each in Extended Data Fig. 1 (also see Methods). Unlike MSGs, which are limited to distinguishing odd and even supercell expansions (via $T\tau$ symmetry) relative to the paramagnetic phase[29], SSGs provide a framework for classifying more complex observed magnetic geometries, especially in cases of noncollinear AFM orders.

In Fig. 1**b**, the spin arrangements of MnTe with the same AFM order but different orientations of the spin moments belong to the same type of SSG $P^{-1}6_3/^{-1}m^{1}m^{-1}c^{\infty m}1$. The SSG preserves all geometric properties of the magnetic structure: its spatial part captures the hexagonal symmetry of the crystal lattice; its spin-only group $^{\infty m}1$ and non-polar $P_{spin}$ $\infty/mm$ accurately describe its collinear AFM character. In comparison, the different MSGs capture the possible physical responses originating from SOC through the coupling of spin and lattice degrees of freedom, which depends on this orientation. Next, we introduce a new theoretical approach, named oriented SSG (OSSG), to unify the SSG and MSG labeling schemes for a comprehensive description of magnetic materials. This formulation enables the SSG, which describes the spin geometry of real materials, to simultaneously account for SOC-related physical effects. More importantly, it provides an intuitive symbol system for visualizing the SOC-induced symmetry-breaking process from SSG to MSG.

***Oriented spin space group (OSSG)***

In the standard International Symbol of SSG, there is no preferred orientation for individual magnetic moments, allowing for a collective *SO*(3) rotation of all magnetic moments by an arbitrary angle. In contrast, a realistic magnetic single crystal typically adopts an energetically favored orientation of its magnetic moments. Therefore, in the OSSG framework, we specify the spin orientation relative to the lattice and implement this feature into the International Symbol system of SSG. As shown in Fig. 1**b**, the OSSG of MnTe is denoted as $P^{-1}6_3/^{-1}m^{1}m^{-1}c^{\infty_{\boldsymbol{n}}m}1$, where the subscript $\boldsymbol{n}$ represents the spin orientation relative to the lattice basis vectors, corresponding to the Néel vector directions [100], [210], and [001] from left to right panels, respectively.

When SOC is imposed, only a subset of the OSSG symmetry operations is preserved, namely those operations $\{g_s||g_l|\tau\}$, where the spatial and spin operations are coupled, so that $g_l$ and $g_s$ contain the same proper rotation. These operations form a subgroup of the OSSG, which can be identified as the MSG of the structure. Consequently, the OSSG assigned to a magnetic structure inherently includes all the symmetry operations of the MSG of the structure as a subgroup, as summarized in Fig. 2**a**. In Fig. 2**b**, we use $Mn_3Sn$ as an example, where the orientation of the spins with respect to the lattice can be specified by the spin direction of the Mn atom with lowest *x* and *z* coordinates. When this orientation is [210], if the spin operations are described in the same basis as the spatial ones the corresponding OSSG can be labelled as $P^{3^{1}_{001}}6_3/^{1}m^{2_{1\text{-}10}}m^{2_{210}}c^{m_{001}}1$, which contains the MSG $Cm'cm'$ (with its standard b axis along [210]). This MSG, which can be generated for instance by the coupled operations $\{m_{001}||m_{001}|0,0,1/2\}$, $\{2_{210}||2_{210}|0,0,1/2\}$ and $\{1||\text{-}1|0\}$ is the one to be assigned to the structure. If instead the orientation is [0-10], the corresponding OSSG is $P^{3^{1}_{001}}6_3/^{1}m^{2_{110}}m^{2_{010}}c^{m_{001}}1$, having a different subgroup as MSG, namely one of type $Cmc'm'$ (with its standard a axis along the [010] direction of the OSSG). This MSG is generated for instance by the coupled operations $\{m_{001}||m_{001}|0,0,1/2\}$,

$\{2_{010}||2_{010}|0\}$ and $\{1||\text{-}1|0\}$.

In essence, given an OSSG, imposing the spin-lattice coupling condition reduces the relevant symmetry to its subgroup formed by all coupled operations. This subgroup constitutes the MSG of the material, under which all possible SOC effects are allowed. From the historical development of magnetism, early studies focused primarily on the MSG, while the application of OSSG was largely overlooked. Incorporating the OSSG perspective can therefore enrich the understanding of the underlying symmetry logic of magnetism. Considering the group-subgroup relation of the OSSG and MSG of a magnetic structure facilitates the study of SOC-driven properties as a symmetry-breaking process. In particular, certain AFM orderings under SOC-related mechanisms, such as Dzyaloshinskii-Moriya (DM) interactions, can exhibit emergent macroscopic magnetization or anomalous Hall effects, which can be rigorously classified as spin-orbit induced magnetism using only symmetry arguments.

***Spin-orbit magnetism (SOM)***

In a well-defined AFM material where zero spin magnetization is enforced by its OSSG, the inclusion of SOC reduces its symmetry to an OSSG subgroup that may no longer enforce $M_s = 0$. It is therefore natural to identify a specific regime within the AFM category based on the comparison of the constraint on the magnetization by the OSSG and by the subgroup symmetry to be considered under spin-lattice coupling. Specifically, the presence of SOC gives rise to a distinct type of magnetism that can be termed spin-orbit magnetism, in which a net spin magnetization arises exclusively from SOC.

As the OSSG subgroup relevant under SOC can be identified with the MSG of the material, we formulate the symmetry criterion for SOM as follows: the material exhibits SOM if its SSG enforces $M_s = 0$, while its MSG does not (see Fig. 3**a**). This criterion unambiguously defines a class of AFM systems that can host FM-like properties related to finite magnetization, such as the AHE[5,30-33], magneto-optical Kerr effect[34,35], etc. More importantly, SOM encompasses previously loosely defined concepts such as weak FM[36-39], while providing a unified symmetry-based theoretical framework for their description. Moreover, SOM also includes recently reported materials exhibiting magnetic-order-induced orbital magnetization[27,28], offering new insights into their relationship with conventional FM and AFM, and the role of SOC.

The OSSG framework offers a comprehensive characterization of the physical properties of SOM, as it inherently captures the pathway of the symmetry breaking induced by SOC. To show how the SOC breaks the OSSG symmetry, we parameterize the SOC Hamiltonian as an SOC tensor, which facilitates tracking its transformation under OSSG operations (see Methods). In this way, one can derive from the constraint of the OSSG whether the SOC-driven magnetization is a first-order or a higher-order effect.

As an example, we apply such theoretical approach to evaluate the spin and orbital magnetizations of noncollinear $Mn_3Sn$ (Fig. 3**b**), which is identified as SOM with an AFM OSSG $P^{3^{1}_{001}}6_3/^{1}m^{2_{110}}m^{2_{010}}c^{m_{001}}1$. The fully coupled operations present in the OSSG, $\{m_{001}||m_{001}|0,0,1/2\}$, $\{2_{010}||2_{010}|0\}$ and $\{1||\text{-}1|0\}$, form the MSG $Cmc'm'$. Under OSSG-enforced constraints for the tensors, both $\boldsymbol{M_s}$ and $\boldsymbol{M_o}$ vectors lie along the OSSG [010] direction (see Methods), as expected from the corresponding MSG. Remarkably, the OSSG allows for separating the possible orbital and spin contributions to the magnetization, which are not distinguished by the MSG. The important information for $Mn_3Sn$ is that the SOC-induced magnetization of orbital origin, $\boldsymbol{M_o}$, is proportional to $\lambda$, whereas $\boldsymbol{M_s}$ is proportional to $\lambda^2$. This different polynomial dependence on SOC strength is confirmed by our density functional theory (DFT) calculations (see Fig. 3**b**). In general, lower-order SOC terms are more significant than higher-order ones, offering additional insights into the magnitude of physical effects. Thus, the OSSG and SOC tensor framework can identify promising AFM candidates with a relatively large AHE (transformed as $\boldsymbol{M_o}$) yet a small net spin magnetization (see Methods).

The symmetry constraints of the magnetizations in FM, SOM, and non-SOM AFM (MSG-enforced $\boldsymbol{M_s} = 0$, termed pure AFM hereafter), summarized in Table I, indicate the fundamental differences among the three types of magnetism. Within the SSG framework, the SOC effects could be described in the *n*th-order term of the SOC expansion, while the non-SOC effects are reflected in the 0th-order term $\lambda^0$. Specifically, the existence of $\lambda^0$ terms in $\boldsymbol{M_s}$ is crucial for distinguishing between the SOM and FM phases, while a pure AFM phase exhibits a clear distinction from SOM and FM ones since in a pure AFM phase both $\boldsymbol{M_s}$ and $\boldsymbol{M_o}$ are zero at all orders of $\lambda$. Furthermore, the OSSG of SOM may allow a nonzero $\boldsymbol{M_o}$, occurring only in the case of noncoplanar materials, as the SSGs of all collinear and coplanar structures force it to be zero[20,21,28,40].

Considering only the MSG of a structure, we can determine whether possible SOC-induced responses, such as AHE, are necessarily zero, but we cannot differentiate magnetic geometries that have the same MSG but a different separation of SOC-free and SOC-driven properties (see Fig. 1**a**). Indeed, the two examples, collinear AFM $CaIrO_3$ and coplanar AFM $Mn_3Ge$, shown in Fig. 1**a**, having the same MSG as the FM $LaMnSi_2$, belong to the SOM category. This can be immediately derived from their OSSG.

***Materials***

The comprehensive symmetry classification introduced in this work provides an efficient route to identify the magnetic order of magnetic materials on a large scale. Such identification has been done using our online program FINDSPINGROUP[41], which can be used to identify the SSG of any given magnetic structure. By this means we have identified the magnetic orders and their OSSG of 2065 experimentally reported magnets available in the MAGNDATA database of the Bilbao Crystallographic Server[24,25]. The dichotomy of FM and AFM by means of their $P_{spin}$ classifies 479 structures as FM materials (including 36 compensated ferrimagnets) and 1586 as AFM materials, accounting for 33.2% and 66.8% of the screened materials, respectively. An exhaustive list of all materials and their OSSG is provided in Supplementary S1.1 and S1.2.

The MSGs of 207 materials identified as AFM (10.0% of total) do not constrain $\boldsymbol{M_s} = 0$, thus corresponding to the SOM category (see Fig. 3**a** and Supplementary S2.1). It is important to stress that the magnetic structures in the MAGNDATA database are experimental structures obtained by neutron diffraction techniques and they may include SOC-driven features, which lower the identified SSG with respect to the one, which would correspond to a virtual SOC-free structure. Consequently, some SOM materials may be wrongly identified as FM because their SOC-induced magnetization exceeds the tolerance threshold used in the SSG assignment. To address this problem, we perform auxiliary evaluations by SOC-free density functional theory (DFT) calculations (see Methods). If the calculated magnetic configuration in the absence of SOC restores a higher-symmetry $P_{spin}$ that constrains $\boldsymbol{M}_s = 0$, the material is reidentified as SOM. This process enabled us to identify 17 additional SOM materials (see Fig. 3**a**, Methods, and Supplementary S2.2).

Finally, we discuss the relationship between recently emergent "new magnetism" and our classification scheme. Our symmetry-based FM/AFM dichotomy, rooted in Néel's original definition, provides the foundation for further categorization of magnetism. The novel magnets, which exhibit zero net magnetization yet unconventional properties, essentially belong to a higher-level classification within the AFM category. For instance, if spin splitting in momentum space (SOC-free) is taken as the target functionality, one can rigorously define spin-split AFM and further classify them into collinear subsets, i.e., altermagnetism, and noncollinear subsets such as *p*-wave magnetism, etc. On the other hand, if SOC-induced magnetization is considered, the corresponding class would be SOM. Therefore, distinct classes of unconventional magnetism do not necessarily encompass one another[11]. For example, spin-split AFM and SOM constitute two independent subclasses of AFM; whereas in 3D collinear magnets, there is a coincidence that SOM form a subset of altermagnets. In the representative altermagnet MnTe, however, the observed AHE does not arise from altermagnetism *per se*, but rather from the fact that its measured magnetic configuration falls within the SOM category. In Extended Data Fig. 3, we illustrate the classifications of the two types of unconventional magnetism.

To summarize, the FM/AFM dichotomy offers a unified conceptual foundation for systematically analyzing and understanding future emergent classifications of magnetism. Within the classification scheme, we introduce the OSSG framework, which provides a natural unification of the classical MSG and the recent SSG descriptions. We recognize that both classical and emerging approaches offer distinct merits for evaluating magnetic materials. Therefore, it is up to the broader magnetism community to evaluate these perspectives and decide which classification approach will be adopted as the standard nomenclature in the future.

Table I. The constraints of symmetry on effects in different magnetic systems. The expansion of SOC is abbreviated as the order form of SOC coefficients. $\lambda^0$ represents the non-SOC effects allowed by OSSG.

| Symmetry | Effects | Pure AFM | SOM | FM |
|---|---|---|---|---|
| OSSG | $M_s$ | 0 | $\sum \lambda^n$ | $\lambda^0 + \sum \lambda^n$ |
| | $M_o$ (AHE) | 0 | $\lambda^0 + \sum \lambda^n$ | $\lambda^0 + \sum \lambda^n$ |
| MSG | $M$ (AHE) | 0 | Not 0 | Not 0 |

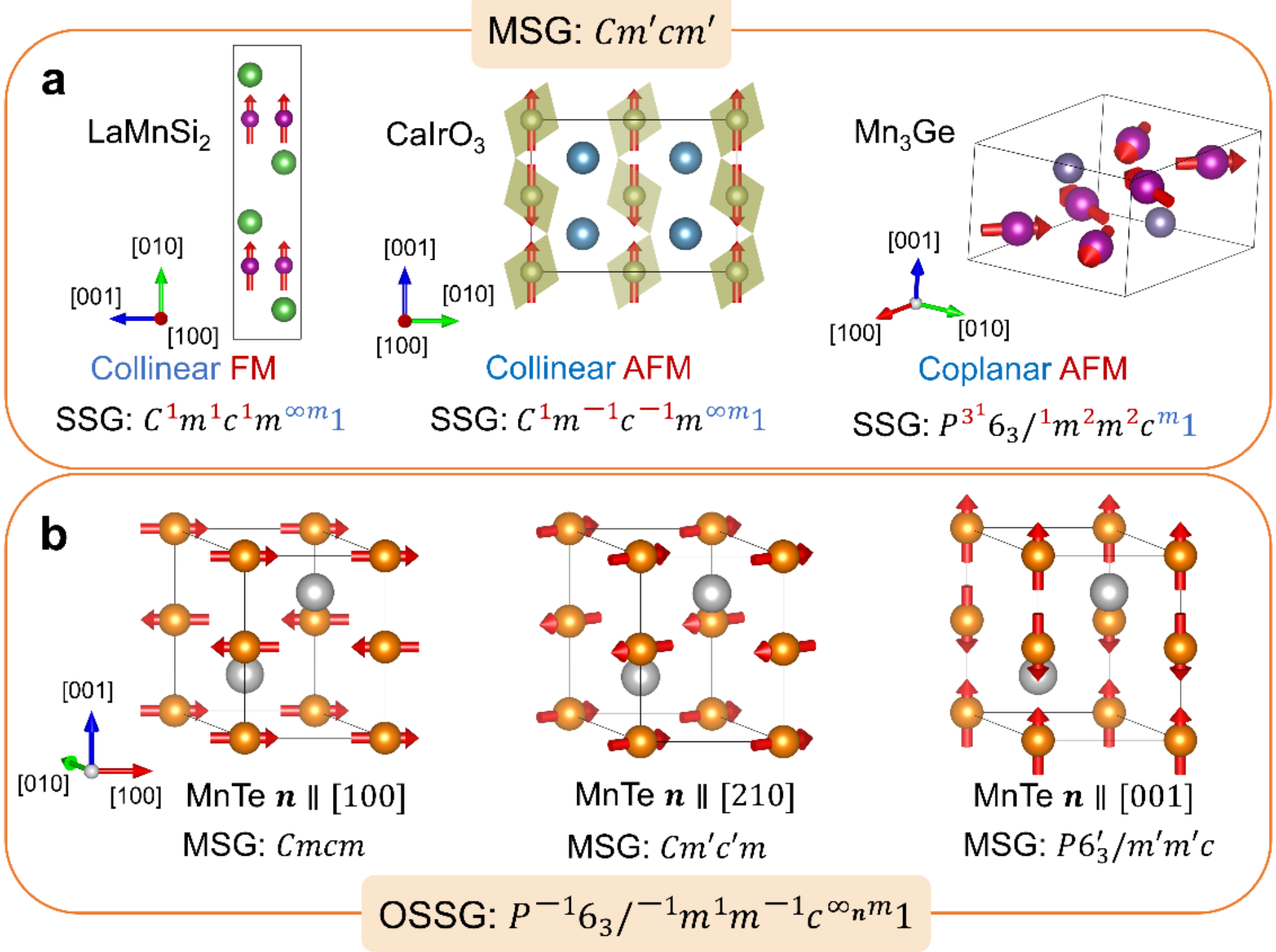

**Fig. 1 | Magnetic structures described by magnetic space groups (MSG) and spin space groups (SSG). a** Distinct magnetic geometries (collinear FM $LaMnSi_2$, collinear AFM $CaIrO_3$ and coplanar AFM $Mn_3Ge$) share the same MSG $Cm'cm'$. These different spin arrangements can be well described by their different SSGs. The components of the SSG symbols that indicate spin dimensionality (collinearity/coplanarity) and magnetic order (FM/AFM) are highlighted in blue and red, respectively. **b** Collinear AFM MnTe with Néel vector $\boldsymbol{n}$ aligned along [100], [210] and [001] crystal orientations gives rise to distinct MSGs but the same SSG, describing its identical collinear AFM arrangement in all cases. In the context of real materials, we have incorporated the information of the MSG into the SSG notation by aligning the basis vectors in real space and spin space, and denoted as oriented SSG (OSSG).

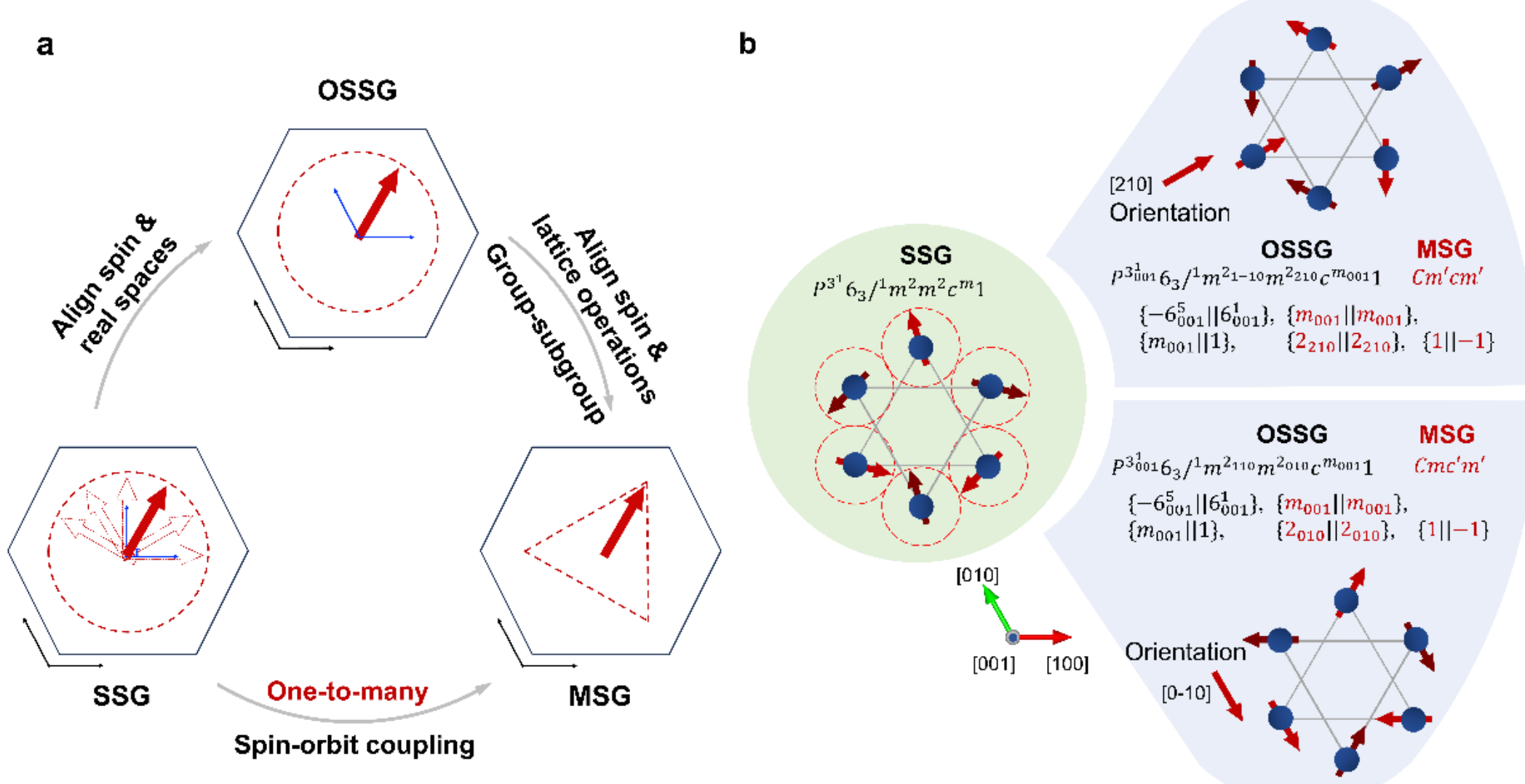

**Fig. 2 | The relation between oriented spin space group (OSSG), spin space group (SSG) and magnetic space group (MSG). a** Since an SSG does not distinguish the collective rotation of local moments (red dashed arrows), considering spin-orbit coupling (SOC) may lead to multiple MSGs. By aligning spin space and real space, an OSSG retains all the operations of an SSG, but its spin-space operations are specified relative to the lattice. Furthermore, SOC aligns spin and lattice operations so that their rotational parts are identical, reducing the OSSG to its subgroup, i.e., the MSG. The red arrow represents a representative vector of the local moment, the black and blue arrows denote the lattice and spin coordinate systems, and the dashed triangle indicates that the MSG is a subgroup obtained from the symmetry breaking of the OSSG. **b** Example of $Mn_3Sn$ with two different spin configurations. In SSG, the spin (red arrows) and lattice (blue spheres and connecting lines) are not coupled, thus the two configurations cannot be distinguished. The OSSG includes all symmetry operations of the SSG but fixes an orientation ([210] or [0-10]) in real space, thereby distinguishing the different configurations. In the MSG, the spin is bound to the lattice, thus contains only those operations for which the real-space and spin-space rotation components are identical (marked in red). For convenience, the translation parts have been omitted.

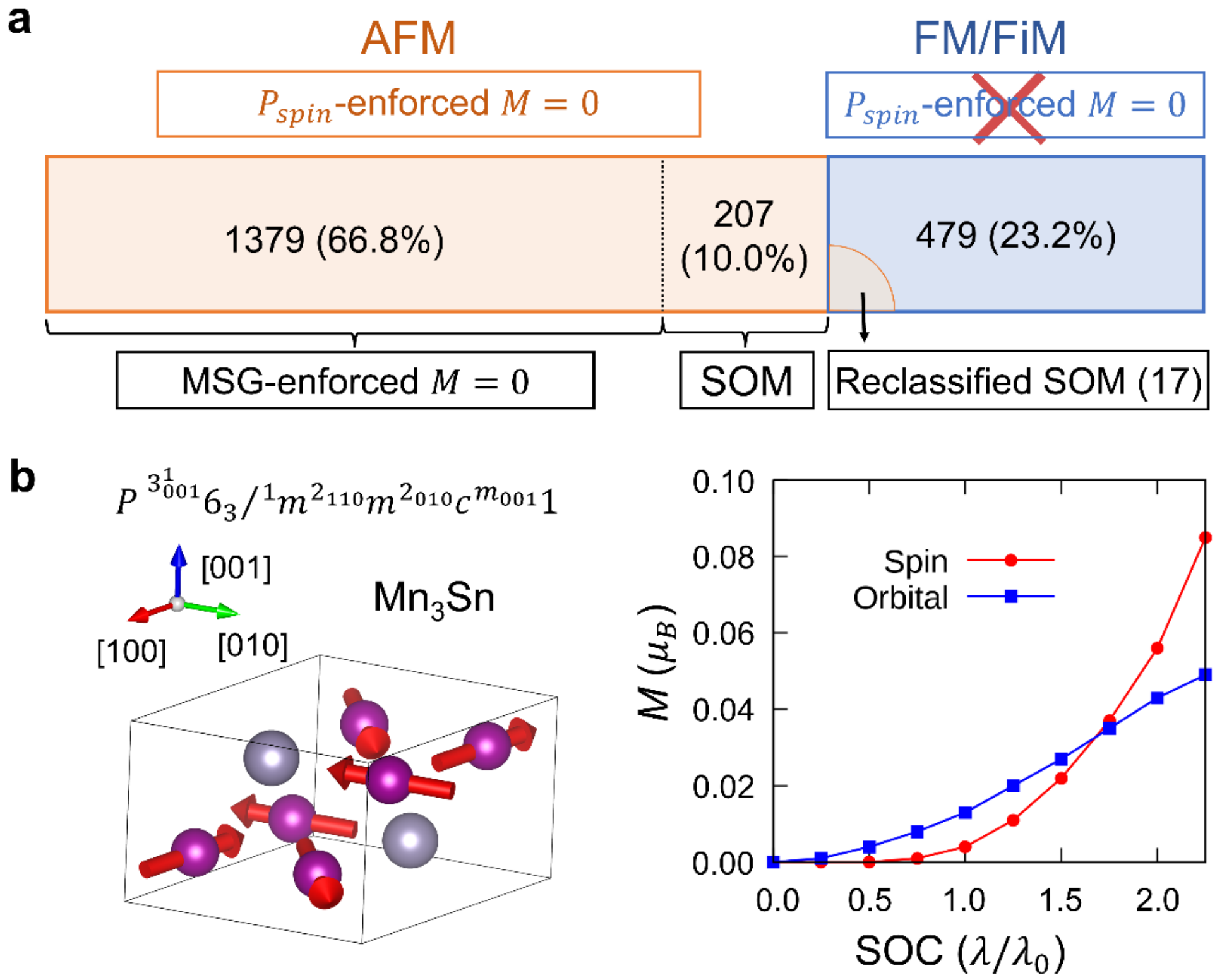

**Fig. 3 | Concept of spin-orbit magnetism (SOM) and classification of magnetic materials. a** Classification of magnetic materials based on the condition of symmetry-enforced $M = 0$ within the spin-space point group $P_{spin}$ [or spin space group (SSG)] and magnetic space group (MSG). The quantities and proportions of each type of materials in the MAGNDATA database are presented, including DFT-assisted identification. **b** Crystal structure and magnetic configuration of $Mn_3Sn$. The spin and orbital magnetizations of $Mn_3Sn$ as a function of spin-orbit coupling (SOC) strength obtained from density functional theory (DFT) calculations. The horizontal axis represents the ratio of the SOC coefficient used in the calculation ($\lambda$) to the actual SOC value for $Mn_3Sn$ ($\lambda_0$).

## Methods

**Further classification of magnetic geometries based on the FM/AFM dichotomy.** Based on the FM/AFM dichotomy, the SSG framework enables further classification of magnetic geometry. Here, we focus on the SSG-based classification of various AFM geometries, especially for noncollinear magnets, that were also phenomenologically described before, such as Néel-type, spiral, and multi-$q$ AFM. Experimentally, the spin distribution across crystallographic primitive cells is typically described by the propagation vector $q$. However, $q$ alone cannot capture the complexity of the magnetic geometry within a single primitive cell. Moreover, when the lattice periodicity and the propagation vector period are mismatched, $q$ fails to reflect the modulation of the crystal field on the magnetic configuration. Furthermore, even the propagation of spiral magnetic order is hardly captured by MSGs, necessitating the application of SSGs.

As mentioned in the main text, we introduce spin translational group $T_{spin}$, which consists of the combination of pure spin space operation and fractional translation $\{g_s||E|\tau\}$. Since the components of $T_{spin}$, $g_s$ and $\tau$, act in different spaces and their multiplicative actions commute, the group $T_{spin}$ follows the group structure of its $\tau$ component and is thus Abelian.

The classification constitutes four distinct categories, as shown in Extended Data Fig. 1. For $i_k$ = 1, $T_{spin}$ only consists of identity operation, and the complexity of magnetic geometry is solely included in the magnetic primary cell. A typical example is CuMnAs with antiparallel spin arrangement for the two Mn atoms within a primary cell. Therefore, such a type of AFM is classified as primary AFM. In the case of $i_k$ = 2, $T_{spin}$ has an order-2 spin translational operation, whose spin-space part can be $-1$, $2$ or $m$. Examples include the intrinsic magnetic topological insulator $MnBi_2Te_4$[42,43], which has two magnetic atoms with antiparallel spin connected by $\{U_2||1|\tau_{1/2}\}$ ($\tau_{1/2} = 0,0,\frac{1}{2}$). Due to the correspondence between the collinear SSG and MSG in the group structure, its group symbol can be simplified as $R_I{}^1\text{-}3^1m^{\infty m}1$. Such a category aligns with the pedagogical one-dimensional AFM chain, referred to as bi-color AFM.

The case of $i_k$ > 2 could be further divided into two categories based on whether

$T_{spin}$ is cyclic. If $T_{spin}$ is a cyclic group, such as $n$, $-n$ $(n > 2)$, the spin geometry aligns with a high-order spin rotation associated with translation. We select $EuIn_2As_2$ as an example that the magnetic moments are connected by $\{U_3||1|\tau_{1/3}\}$, forming a so-called spiral AFM[21,44]. Finally, if $T_{spin}$ is a non-cyclic Abelian group, the spin rotations with different axes must be mapped to translations in different directions. Such mappings result in a more intricate multiple-$q$ spin geometry, as observed in AFM [111]-strained cubic γ-FeMn[27] and $CoNb_3S_6$[28], referred to as multi-axial AFM. Apparently, both spiral and multi-axial AFM cannot be described by MSGs, in which the corresponding $T_{spin}$ only allows $\{\text{-}1||1|\tau\}$ operation. Furthermore, FM can also be classified into the four categories following the same way. For example, a helimagnet with AFM geometries and an FM magnetic canting can be directly described by combining a $T_{spin}$ with $i_k > 2$ and a polar $P_{spin}$.

Extended Data Fig. 2 summarizes the quantities and proportions of materials exhibiting each type of AFM geometry in the MAGNDATA database obtained by our online program FINDSPINGROUP. Based on $T_{spin}$, AFM geometries are further classified into primary (660, 32.0%), bi-color (857, 41.5%), spiral (24, 1.2%) and multi-axial (45, 2.2%) categories. In Supplementary S2.1 and S2.2, we provide an exhaustive list of all materials and their oriented SSG including the dichotomy of FM/AFM and further geometries classification based on $T_{spin}$.

**Spin-orbit coupling (SOC) tensor.** To describe the transformation of SOC under SSG operations, we reformulate it in a form that explicitly allows for independent coordinate systems in real space and spin space:

$$\hat{H}_{SOC} = \lambda \hat{\boldsymbol{L}}^T \boldsymbol{\chi} \hat{\boldsymbol{\sigma}} = \lambda \sum_{i,j} \chi_{ij} \hat{L}_i \hat{\sigma}_j$$

$$= \lambda (\hat{L}_1 \quad \hat{L}_2 \quad \hat{L}_3) \begin{pmatrix} \boldsymbol{r}_1 \cdot \boldsymbol{s}_1 & \boldsymbol{r}_1 \cdot \boldsymbol{s}_2 & \boldsymbol{r}_1 \cdot \boldsymbol{s}_3 \\ \boldsymbol{r}_2 \cdot \boldsymbol{s}_1 & \boldsymbol{r}_2 \cdot \boldsymbol{s}_2 & \boldsymbol{r}_2 \cdot \boldsymbol{s}_3 \\ \boldsymbol{r}_3 \cdot \boldsymbol{s}_1 & \boldsymbol{r}_3 \cdot \boldsymbol{s}_2 & \boldsymbol{r}_3 \cdot \boldsymbol{s}_3 \end{pmatrix} \begin{pmatrix} \hat{\sigma}_1 \\ \hat{\sigma}_2 \\ \hat{\sigma}_3 \end{pmatrix}, \quad (1)$$

where $\lambda$, $\hat{\boldsymbol{L}}$ and $\hat{\boldsymbol{\sigma}}$ represent the SOC coefficient, effective orbital angular momentum operator and spin operator, respectively; $\boldsymbol{r}_i$ and $\boldsymbol{s}_j$ are the unit base vectors with $i = 1,2,3$ and $j = 1,2,3$ for real space and spin space, respectively; $\boldsymbol{\chi}$ represents a $3 \times 3$ SOC tensor matrix, defined as $\boldsymbol{\chi} = \{\chi_{ij} = \boldsymbol{r}_i \cdot \boldsymbol{s}_j | i = 1,2,3; j = 1,2,3\}$. For a general SSG operation $\{g_s||g_l\}$, the transformation of $\boldsymbol{\chi}$ can be expressed as:

$$\hat{\rho}^{-1}_{\{g_s||g_l\}} \lambda \hat{\boldsymbol{L}}^T \boldsymbol{\chi} \hat{\boldsymbol{\sigma}} \, \hat{\rho}_{\{g_s||g_l\}}$$

$$= \lambda det(R_s) det(R_l) \left[\hat{\boldsymbol{L}}^T R_l^{-1}\right] \boldsymbol{\chi} [R_s \hat{\boldsymbol{\sigma}}]$$

$$= \lambda \hat{\boldsymbol{L}}^T det(R_s) det(R_l) [R_l^{-1} \boldsymbol{\chi} R_s] \hat{\boldsymbol{\sigma}}, \quad (2)$$

where $\hat{\rho}_{\{g_s||g_l\}}$ is the representation operator of $\{g_s||g_l\}$ in Hilbert space; $R_l$ and $R_s$ are 3D Euclidean transformations corresponding to $g_l$ and $g_s$ in three-dimensional real space and spin space, respectively. $det(R_l)$ and $det(R_s)$ are the determinants of $R_l$ and $R_s$, respectively; their values, either -1 or 1, depend on whether $R_l$ includes space-inversion operation and whether $R_s$ includes time-reversal operation, respectively. Therefore, the transformation of the SOC term under a SSG operation can be described using the SOC tensor $\boldsymbol{\chi}$, based on its defined transformation rule:

$$\boldsymbol{\chi} \xrightarrow{\{g_s||g_l\}} \boldsymbol{\chi}' = det(R_s) det(R_l) R_l^{-1} \boldsymbol{\chi} R_s. \quad (3)$$

A similar method has also been applied to investigate the anomalous Hall effect in ferromagnetic systems[45].

**Material example for orbital and spin magnetization: $Mn_3Sn$.** In the following, we use the orbital magnetization $\boldsymbol{M}_o$, the spin magnetization $\boldsymbol{M}_s$ and noncollinear AFM $Mn_3Sn$ (Fig. 3b) as examples to demonstrate how to analyze the SOC-induced physical properties by SOC tensor $\boldsymbol{\chi}$. By definition, $\boldsymbol{M}_o$ and $\boldsymbol{M}_s$ are the sums of the expectation values of orbital angular momentum operator $\hat{\boldsymbol{\mathcal{L}}}$ and spin operator $\hat{\boldsymbol{\sigma}}$ over the entire Brillouin zone, respectively, expressed as:

$$\boldsymbol{M}_o = -\frac{\mu_B g_o}{2\pi} \int_{BZ} \sum_n f_{n\boldsymbol{k}} \langle \varphi_n(\boldsymbol{k}) | \hat{\boldsymbol{\mathcal{L}}} | \varphi_n(\boldsymbol{k}) \rangle \, d\boldsymbol{k}, \quad (4)$$

$$\boldsymbol{M}_s = -\frac{\mu_B g_s}{2\pi} \int_{BZ} \sum_n f_{n\boldsymbol{k}} \langle \varphi_n(\boldsymbol{k}) | \hat{\boldsymbol{\sigma}} | \varphi_n(\boldsymbol{k}) \rangle \, d\boldsymbol{k}, \quad (5)$$

where $f_{n\boldsymbol{k}}$ is the Fermi distribution at wave vector $\boldsymbol{k}$; $\mu_B$ represents the Bohr magneton; $g_o$ and $g_s$ denote the Landé $g$-factors for orbital and spin, respectively. Consequently, the transformations of $\boldsymbol{M}_o$ and $\boldsymbol{M}_s$ are equivalent to time-reversal-odd axial vectors, and follow the corresponding proper rotation operations in real space and spin space, respectively, expressed as:

$$\boldsymbol{M}_o \xrightarrow{\{g_s||g_l\}} det(R_s)det(R_l)R_l\boldsymbol{M}_o, \tag{6}$$

$$\boldsymbol{M}_s \xrightarrow{\{g_s||g_l\}} R_s\boldsymbol{M}_s. \tag{7}$$

To analyze the coupling relationship between $\boldsymbol{M}_o$, $\boldsymbol{M}_s$ and $\lambda\boldsymbol{\chi}$, we express both $\boldsymbol{M}_o$ and $\boldsymbol{M}_s$ as a series expansion in terms of $\lambda\boldsymbol{\chi}$:

$$M_a[\boldsymbol{\chi}] = \omega_a^{(0)} + \lambda\sum_{ij}\omega_{a,ij}^{(1)}\chi_{ij} + \lambda^2\sum_{ijkl}\omega_{a,ij,kl}^{(2)}\chi_{ij}\chi_{kl} + \cdots, \tag{8}$$

where $\boldsymbol{\omega}^{(n)}$ is a $(2n+1)$th-order undetermined tensor. Equation (8) is constrained by the OSSG symmetry and is valid both for $\boldsymbol{M}_o$ and $\boldsymbol{M}_s$, but in each case different transformation properties have to be considered for the tensors $\boldsymbol{\omega}^{(n)}$ due to the restriction:

$$det(R_s)det(R_l)R_l\,\boldsymbol{M}_o[\boldsymbol{\chi}] = \boldsymbol{M}_o[det(R_s)det(R_l)R_l\boldsymbol{\chi}R_s^{-1}], \tag{9}$$

$$R_s\boldsymbol{M}_s[\boldsymbol{\chi}] = \boldsymbol{M}_s[det(R_s)det(R_l)R_l\boldsymbol{\chi}R_s^{-1}]. \tag{10}$$

Once the properties of $\boldsymbol{\omega}^{(n)}$ have been established, the same basis in real and spin space can be chosen (i.e., $\chi_{ij} = \delta_{ij}$), fixing the spin orientation to that defined by the OSSG. Equation (8) then strongly simplifies and only very specific components of the OSSG symmetry-adapted tensors $\boldsymbol{\omega}^{(n)}$ become relevant.

The SOM material $Mn_3Sn$ has the OSSG $P^{3_{001}^1}6_3/^1m^{2_{110}}m^{2_{010}}c^{m_{001}}1$. The point operation parts of the OSSG generators include $\{1||\text{-}1\}$, $\{\text{-}6_{001}^5||6_{001}^1\}$, $\{2_{100}||2_{1\text{-}10}\}$ and $\{m_{001}||1\}$. By applying the symmetry constraints of all the OSSG generators and choosing $\chi_{ij} = \delta_{ij}$, we can analyze the relationship between $\boldsymbol{M}_o$, $\boldsymbol{M}_s$ and $\boldsymbol{\chi}$ order by order. For the zeroth-order SOC tensor term, the coefficient $\omega^{(0)}$ remains invariant under all OSSG operations. However, both $\boldsymbol{M}_o$ and $\boldsymbol{M}_s$ transform non-identity under this OSSG. Therefore, the zeroth-order $\omega^{(0)}$ must vanish for all three components of $\boldsymbol{M}_o$ and $\boldsymbol{M}_s$. For the first-order SOC tensor term, by requiring that each component of the tensor $\boldsymbol{\omega}^{(1)}$ remains invariant under the OSSG generators, the OSSG restriction on the first-order coefficient tensor $\boldsymbol{\omega}_o^{(1)}$ for $\boldsymbol{M}_o$ can be obtained by combining equations (8) and (9). Applying the same method to the symmetry constraints of $\boldsymbol{M}_s$, we find that all first-order SOC tensor terms of $\boldsymbol{M}_s$ are forbidden by the SSG symmetry. Expressed in an orthonormal basis parallel to the directions $(\boldsymbol{a}, 2\boldsymbol{b}+\boldsymbol{a}, \boldsymbol{c})$, the expansion of the spin magnetization $\boldsymbol{M}_s$ and the orbital magnetization $\boldsymbol{M}_o$ (to the lowest non-zero order terms) can be written as:

$$M_{s,1} = 2\left(\omega_{s,1,1111}^{(2)} + \omega_{s,1,2222}^{(2)}\right)\lambda^2, \quad M_{s,2} = -2\sqrt{3}\left(\omega_{s,1,1111}^{(2)} + \omega_{s,1,2222}^{(2)}\right)\lambda^2, \tag{11}$$

$$M_{o,1} = 2\omega_{o,1,11}^{(1)}\lambda, \qquad M_{o,2} = -2\sqrt{3}\omega_{o,1,11}^{(1)}\lambda, \tag{12}$$

where $M_{s,i}$ and $M_{o,i}$ represent the $i$ th components of $\boldsymbol{M}_s$ and $\boldsymbol{M}_o$ in the mentioned basis, respectively. Both $\boldsymbol{M}_s$ and $\boldsymbol{M}_o$ vectors lie therefore along the OSSG [010] direction, as expected from the corresponding MSG. Note, however, that equations (11) and (12) have been obtained by applying the symmetry conditions of the OSSG, with no explicit use of the MSG.

Next, we discuss the advantages of the SOC tensor framework in identifying promising AFM candidates for the anomalous Hall effect (AHE). According to the Kubo formula, the intrinsic anomalous Hall conductivity can be expressed as:

$$\sigma_z^{AHE} = \frac{e^2}{\hbar}\sum_{n'\neq n}\int_{BZ}\frac{d^3k}{(2\pi)^3} f_{\boldsymbol{k}n}\frac{2Im\left[\langle\boldsymbol{k}n|\partial_{k_x}\hat{H}(\boldsymbol{k})|\boldsymbol{k}n'\rangle\langle\boldsymbol{k}n'|\partial_{k_y}\hat{H}(\boldsymbol{k})|\boldsymbol{k}n\rangle\right]}{\left(\epsilon_{\boldsymbol{k}n}-\epsilon_{\boldsymbol{k}n'}\right)^2}, \tag{13}$$

where $f_{\boldsymbol{k}n}$ is the Fermi-Dirac distribution, $\hat{H}(\boldsymbol{k})$ is the system Hamiltonian, and $|\boldsymbol{k}n\rangle$, $|\boldsymbol{k}n'\rangle$ are the eigenstates of the system. From symmetry considerations, the anomalous Hall conductivity vector $\boldsymbol{\sigma}^{AHE} = (\sigma_x^{AHE}, \sigma_y^{AHE}, \sigma_z^{AHE})$ transforms as an axial vector in real space and is odd under time-reversal symmetry in spin space. Therefore, it shares the same symmetry transformation properties as the orbital magnetic moment $\boldsymbol{M}_o$, and consequently follows the same expansion form in terms of the SOC tensor. Within the MSG framework that includes SOC, AHE and magnetization are subject to the same symmetry constraints—meaning that symmetry either permits both or forbids both simultaneously. On the other hand, our SOC tensor framework enables a systematic comparison of the magnitudes of the AHE

(transformed as $\boldsymbol{M}_o$) and the spin magnetization $\boldsymbol{M}_s$, providing insights for realizing a large AHE response in systems with minimal net magnetization.

**Identification of spin-orbit magnetism (SOM) materials.** According to the symmetry classification in our paper, SOMs exhibit SSG-enforced $M_s = 0$ but not MSG-enforced $M = 0$, indicating that the net magnetization originates from SOC. To identify the SOM materials in the MAGNDATA database[24,25], we use the FINDSPINGROUP program[41] to identify the SSG and MSG of all the materials with tolerance $\Delta = 0.02\ \mu_B$, and find 207 SOM materials (left workflow in Extended Data Fig. 4). Here, the tolerance $\Delta$ is defined as the allowable magnitude of the vector difference $|\boldsymbol{M}_i - R_s\boldsymbol{M}_j|$, where $\boldsymbol{M}_i$ and $\boldsymbol{M}_j$ are the magnetic moments at atomic sites $i$ and $j$, respectively, satisfying the mappings $j \overset{g_l}{\rightarrow} i$ and $\boldsymbol{M}_j \overset{g_s}{\rightarrow} R_s\boldsymbol{M}_j$ under the symmetry operation $\{g_s||g_l\}$; $R_s$ is a 3D orthogonal transformation in spin space corresponding to $g_s$.

Next, to distinguish materials in which the net magnetization is generated by SOC but SSG has been identified as FM, we increase the tolerance to 1.50 $\mu_B$, resulting in the symmetry identification of 61 additional possible SOM materials. After excluding materials with strong disorder, using SOC-free DFT calculations, we compare for each material the energy of the magnetic arrangement provided by MAGNDATA with arrangements having a higher SSG. The results indicate that the SOC-free ground states of 17 materials have a SSG of higher symmetry, which does not allow $M_s$, and are re-identified as SOM materials, with their $M_s$ being characterized as a SOC-driven effect (right workflow in Extended Data Fig. 4). The remaining 44 materials are mainly FM systems with tiny magnetizations and disordered systems that are beyond the scope of this symmetry identification. The list of SOMs is provided in Supplementary S2.1 and the DFT reidentified SOM materials are provided in Supplementary S2.2.

In the following, we present material examples to demonstrate the identification of SOM materials. For direct symmetry identification, we use $LaMnO_3$ as an example, where the three components of the local magnetic moments are (3.87 $\mu_B$, 0.0, 0.0). Although the symmetry constraint of its net magnetic moment is (0, $M_y$, 0), the local magnetic moment does not have a $y$-direction component within the accuracy (0.02 $\mu_B$) allowed by the database. As a result, the FINDSPINGROUP program can directly identify its magnetic geometry of collinear AFM and classify it as SOM. The SOM materials with experimental negligible net magnetic moment account for 10.0% of the entire material database, which is the most common situation in SOM.

On the other hand, some materials with SOC induced net magnetic moments are sufficiently large, requiring auxiliary evaluation via DFT calculations. We use $NiF_2$ as an example, where the three components of the local magnetic moments are (2.0 $\mu_B$, 0.03 $\mu_B$, 0.0). As a result, its net magnetic moment in a unit cell is 0.06 $\mu_B$, which requires evaluation to determine whether it origiates from SOC. We perform SOC-free DFT calculations to compare the total energy of this magnetic configuration with that of the configuration without canting. The results show that the magnetic configuration without canting has lower energy, indicating that the net magnetic moment is induced by SOC. Furthermore, we re-identify the symmetry of the magnetic configuration without canting to confirm the OSSG of the ground-state magnetic configuration without SOC. The revised-OSSG of $NiF_2$ is $P^{-1}4_2/^{1}m^{-1}n^{1}m^{\infty_{100}m}1$, confirming that it is SOM.

**Density functional theory (DFT) calculations.** Our DFT calculations are conducted using the Vienna ab initio simulation package (VASP)[46] that employed the projector augmented wave (PAW)[47] method. The exchange-correlation functional was described through the generalized gradient approximation of the Perdew-Burke-Ernzerhof formalism (PBE)[48,49] with on-site Coulomb interaction Hubbard U, which are provided in Supplementary S2.2 for each material. The plane-wave cutoff energy was set to 500 eV, and total energy convergence criteria was set to $1.0\times10^{-6}$ eV for all candidate materials. Sampling of the entire Brillouin zone was performed by a Γ-centered Monkhorst-Pack grid[50], with the standard requiring that the product of the number of *k*-points and the lattice constant exceeds 45 Å for each direction.

## Method References

**Acknowledgments**

We thank J. Liu and Y. Gao for the helpful discussions. This work was supported by National Natural Science Foundation of China under Grant No. 12525410, No. 12274194, No. 12574275 and No.12534003, the National Key R&D Program of China under grant no. 2025YFA1411300, Guangdong Provincial Quantum Science Strategic Initiative under Grant No. GDZX2401002, Guangdong Provincial Key Laboratory for Computational Science and Material Design under Grant No. 2019B030301001, Shenzhen Science and Technology Program (Grant No. RCJC20221008092722009 and No. 20231117091158001), the Innovative Team of General Higher Educational Institutes in Guangdong Province (Grant No. 2020KCXTD001), the Open Fund of the State Key Laboratory of Spintronics Devices and Technologies (Grant No. SPL-2407) and Center for Computational Science and Engineering of Southern University of Science and Technology.

**Author contributions**

Q.L. conceived the project. Y.L., X.C. and Q.L. constructed the magnetic classification. Y.L., X.C., J.M.P.-M. and Q.L. constructed the theory of oriented spin space group. Y.L., J.E. and J.M.P.-M. derived the spin-orbit tensor formalism. Y.Y. deployed the identification of OSSG in FINDSPINGROUP and performed the screening of SOM materials. Y.L., X.C. and Y.Y. performed the DFT calculations and established the database. Y.L., X.C. and Q.L. analyzed the results and wrote the paper with the input of all the authors.

**Competing interests**

The authors declare no competing interests.

**Additional information**

**Supplementary information** The online version contains supplementary material available at https://xxxx.

**Correspondence and requests for materials** should be addressed to Qihang Liu.

## Data availability

All data are available in the Supplementary Information and through our public website, the online program for identifying the spin (magnetic) space group symmetries and related properties of magnetic materials (https://findspingroup.com/).

## Code availability

All codes are available through our public website, the online program for identifying the spin (magnetic) space group symmetries and related properties of magnetic materials (https://findspingroup.com/).

**Extended Data Table I.** All possible spin-space point groups ($P_{spin}$) classified by group structures and the corresponding FM/AFM dichotomy. The subscript $n$ represents the helicity of the magnetic moments, and $\infty$ indicates that the system gains spin-space $U(1)$ symmetry, corresponding to collinear magnetic structures. International Symbols are used to present the point groups.

| $P_{spin}$ | Magnetism |
|---|---|
| $n$ | FM |
| $nmm, nm, \infty m$ | FM |
| -$n, n/m$ | AFM |
| $n22, n2$ | AFM |
| -$nm$,-$n2m, n/mmm, \infty/mm$ | AFM |
| $cubic$ ($23, m$-3,432,-43$m, m$-3$m$) | AFM |

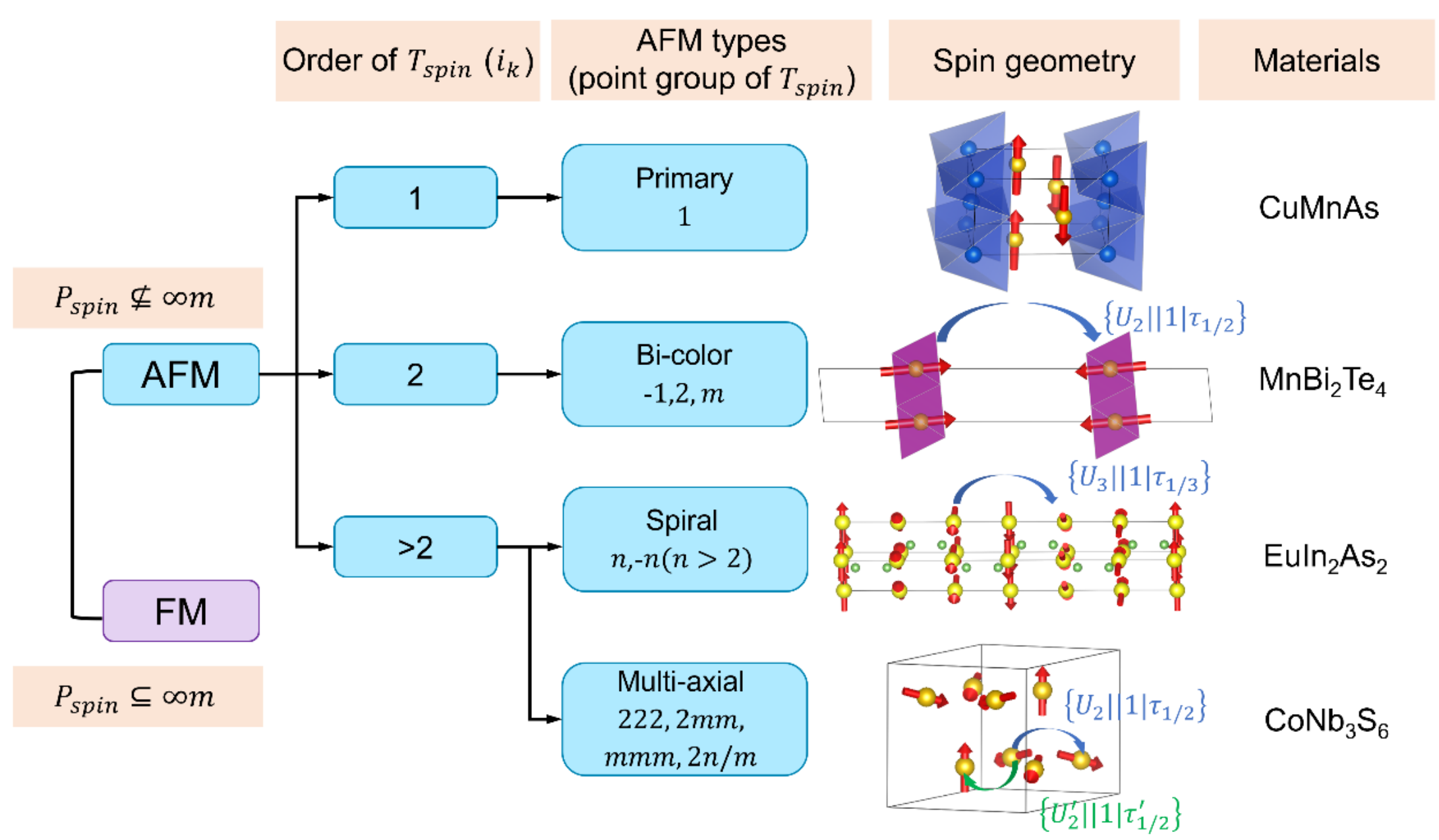

**Extended Data Fig. 1 | Classification of magnetic orders.** The dichotomy of FM and AFM orders is classified by the spin-space point group $P_{spin}$. Furthermore, the geometries of AFM are classified into four categories, i.e., primary, bi-color, spiral, and multi-axial, based on the order of the spin translational group $T_{spin}$ and whether it forms a cyclic group (for spiral and multi-axial categories). Examples of representative materials with their magnetic geometries (only magnetic ions are shown) are provided. The spin translational operations of the corresponding SSG for each material are also denoted, where $U_n$ stands for a *n*-fold rotation in spin space.

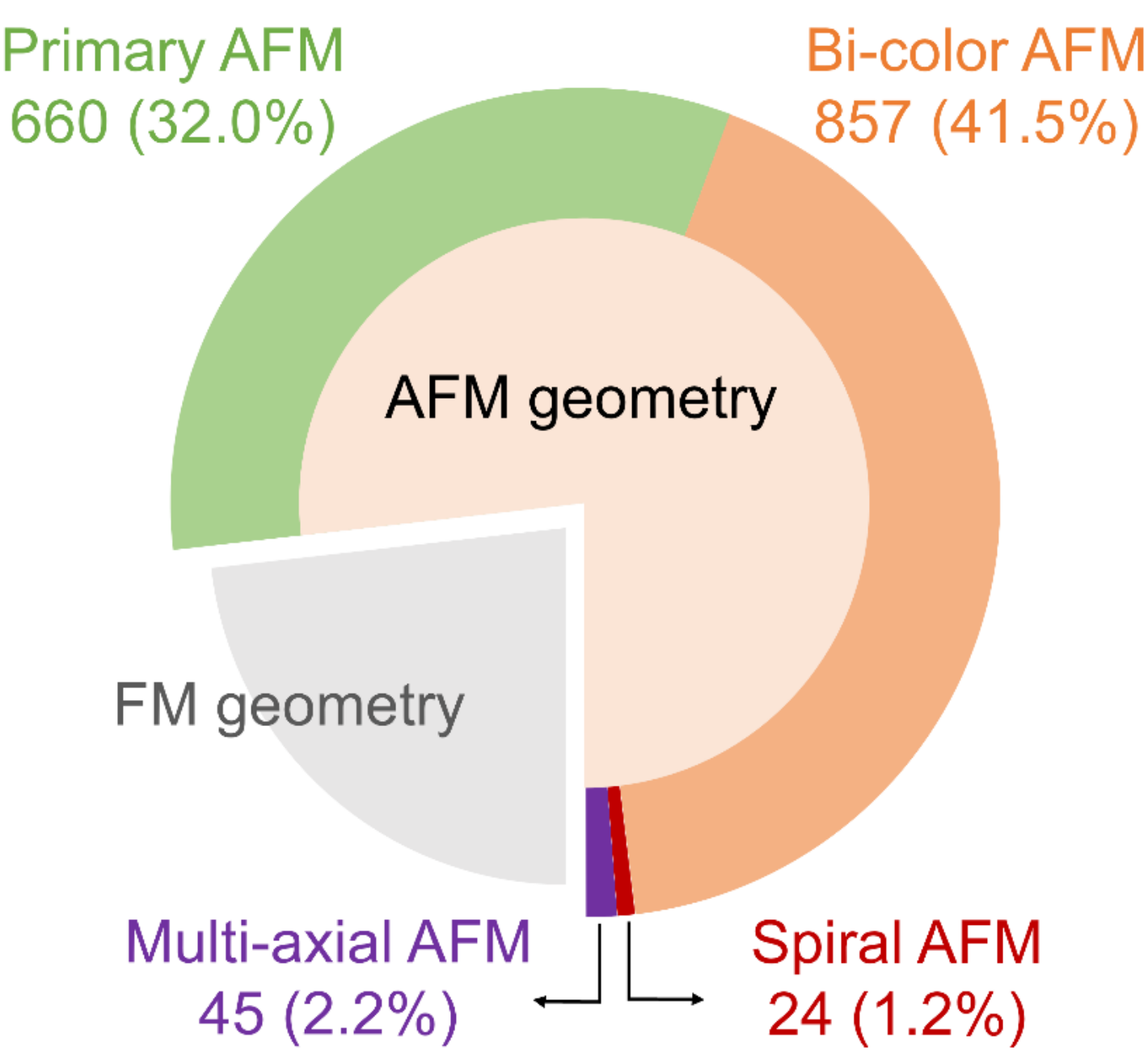

**Extended Data Fig. 2 | The quantities and proportions of various AFM geometries in the MAGNDATA database.**

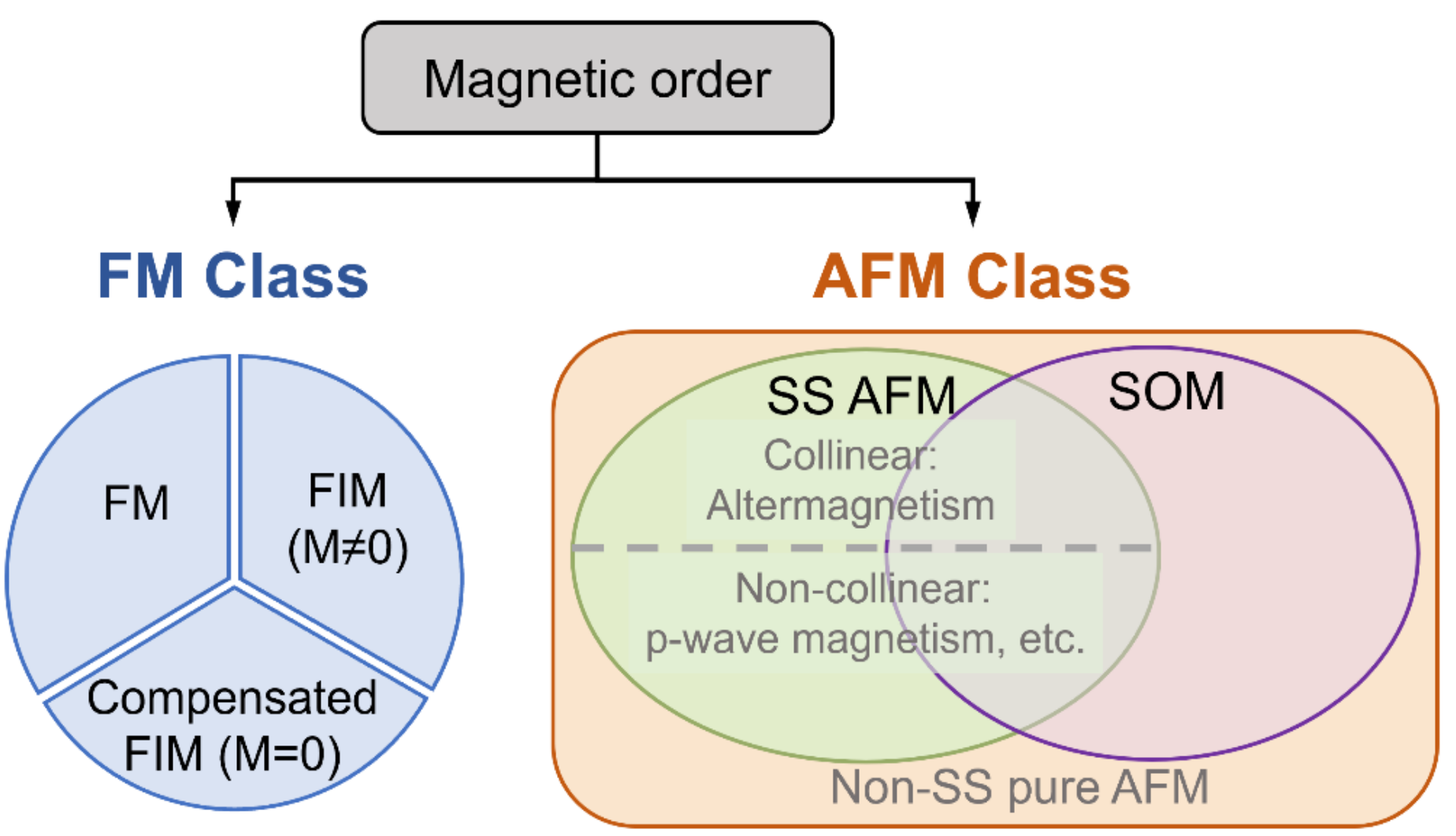

**Extended Data Fig. 3 | Relationship between ferromagnetism (FM) / antiferromagnetism (AFM) dichotomy and further classification schemes, including altermagnetism and spin-orbit magnetism (SOM).** The FM class can be further classified into ferromagetism ($M \neq 0$), ferrimagnetism (FIM, $M \neq 0$), and compensated ferrimagnetism ($M = 0$). The AFM class can also be further classified based on different properties as criteria. Therefore, distinct classes of corresponding classification do not necessarily encompass each other. Non-SS pure AFM stands for AFM without either spin splitting or SOC-induced magnetization.

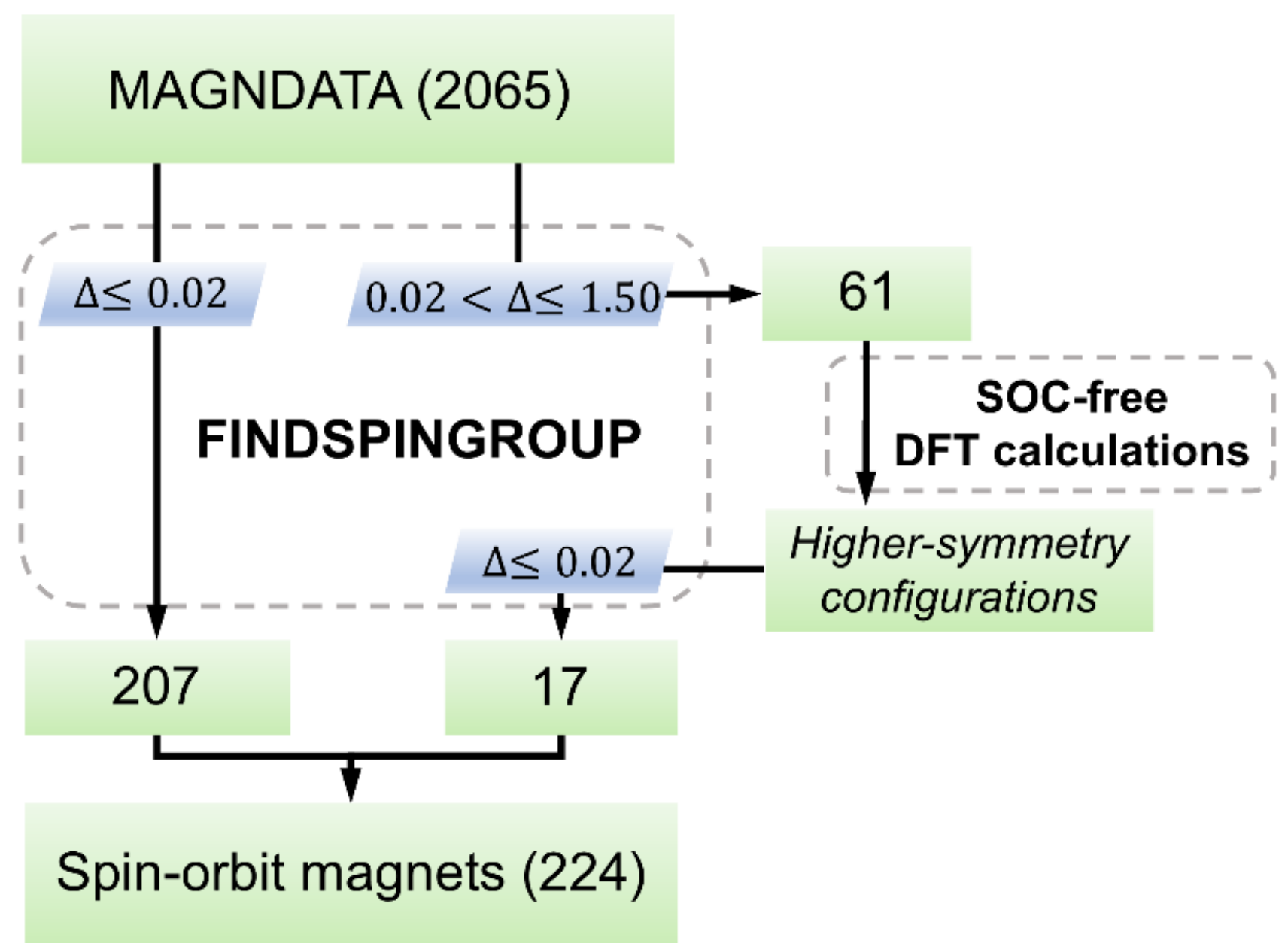

**Extended Data Fig. 4 | Workflow of the identification of spin-orbit magnetism (SOM).** The left workflow shows the direct identification of SOM using the FINDSPINGROUP program, while the right workflow shows the identification process assisted by SOC-free density DFT calculations. The numbers in the green rectangular boxes represent the quantity of materials obtained at each step. The blue diagonal boxes indicate the tolerance range used by FINDSPINGROUP.